\documentclass[conference]{IEEEtran}
\IEEEoverridecommandlockouts
\usepackage{cite}
\usepackage{amsmath,amssymb,amsfonts}
\usepackage{algorithmic}
\usepackage{graphicx}
\usepackage{textcomp}
\usepackage{xcolor}
\usepackage[hidelinks]{hyperref}
\usepackage{comment}

\def\BibTeX{{\rm B\kern-.05em{\sc i\kern-.025em b}\kern-.08em
    T\kern-.1667em\lower.7ex\hbox{E}\kern-.125emX}}

\begin{document}

\title{cuHPX: GPU-Accelerated Differentiable Spherical Harmonic Transforms on HEALPix Grids\\
}

\author{\IEEEauthorblockN{Xiaopo Cheng}
\IEEEauthorblockA{
\textit{NVIDIA}\\
Santa Clara, USA \\
xiaopoc@nvidia.com}
\and
\IEEEauthorblockN{Akshay Subramaniam* \thanks{* Corresponding author: Akshay Subramaniam, NVIDIA, 2788 San Tomas Expy, Santa Clara, CA 95051}}
\IEEEauthorblockA{
\textit{NVIDIA}\\
Santa Clara, USA \\
asubramaniam@nvidia.com}
\and
\IEEEauthorblockN{Shixun Wu}
\IEEEauthorblockA{\textit{NVIDIA}\\
Santa Clara, USA \\
swu264@ucr.edu}
\and
\IEEEauthorblockN{Noah Brenowitz}
\IEEEauthorblockA{\textit{NVIDIA}\\
Santa Clara, USA\\
nbrenowitz@nvidia.com}
}

\maketitle

\begin{abstract}
HEALPix (Hierarchical Equal Area isoLatitude Pixelization) is a widely adopted spherical grid system in astrophysics, cosmology, and Earth sciences. Its equal-area, iso-latitude structure makes it particularly well-suited for large-scale data analysis on the sphere. However, implementing high-performance spherical harmonic transforms (SHTs) on HEALPix grids remains challenging due to irregular pixel geometry, latitude-dependent alignments, and the demands for high-resolution transforms at scale. In this work, we present \texttt{cuHPX}, an optimized CUDA library that provides functionality for spherical harmonic analysis and related utilities on HEALPix grids. Beyond delivering substantial performance improvements, \texttt{cuHPX} ensures high numerical accuracy, analytic gradients for integration with deep learning frameworks, out-of-core memory-efficient optimization, and flexible regridding between HEALPix, equiangular, and other common spherical grid formats. Through evaluation, we show that \texttt{cuHPX} achieves rapid spectral convergence and delivers \textbf{over 20$\times$ speedup} compared to existing libraries, while maintaining numerical consistency. By combining accuracy, scalability, and differentiability, \texttt{cuHPX} enables a broad range of applications in climate science, astrophysics, and machine learning, effectively bridging optimized GPU kernels with scientific workflows.

Code is available at \href{https://github.com/NVlabs/cuHPX}{https://github.com/NVlabs/cuHPX}.
\end{abstract}


\section{Introduction}

Analyzing massive spherical data sets, ranging from global climate simulations to cosmic microwave background (CMB) measurements, requires efficient methods to represent and process functions defined on the sphere. Spherical Harmonic Transforms (SHTs) provide a fundamental mathematical tool for this purpose, enabling the decomposition of spatial data into spectral components for compact representation and large-scale analysis. However, the accuracy and efficiency of SHTs depend critically on the choice of spherical grid. A natural discretization is the equiangular latitude–longitude grid, which partitions the sphere using evenly spaced parallels and meridians. While conceptually simple, this representation produces pixels of highly varying size: near the poles, pixels are more densely populated and distorted compared to those near the equator. These irregularities degrade both the numerical stability and computational efficiency of large-scale SHTs.

\begin{figure}[h]
    \centering
    \includegraphics[width=\linewidth]{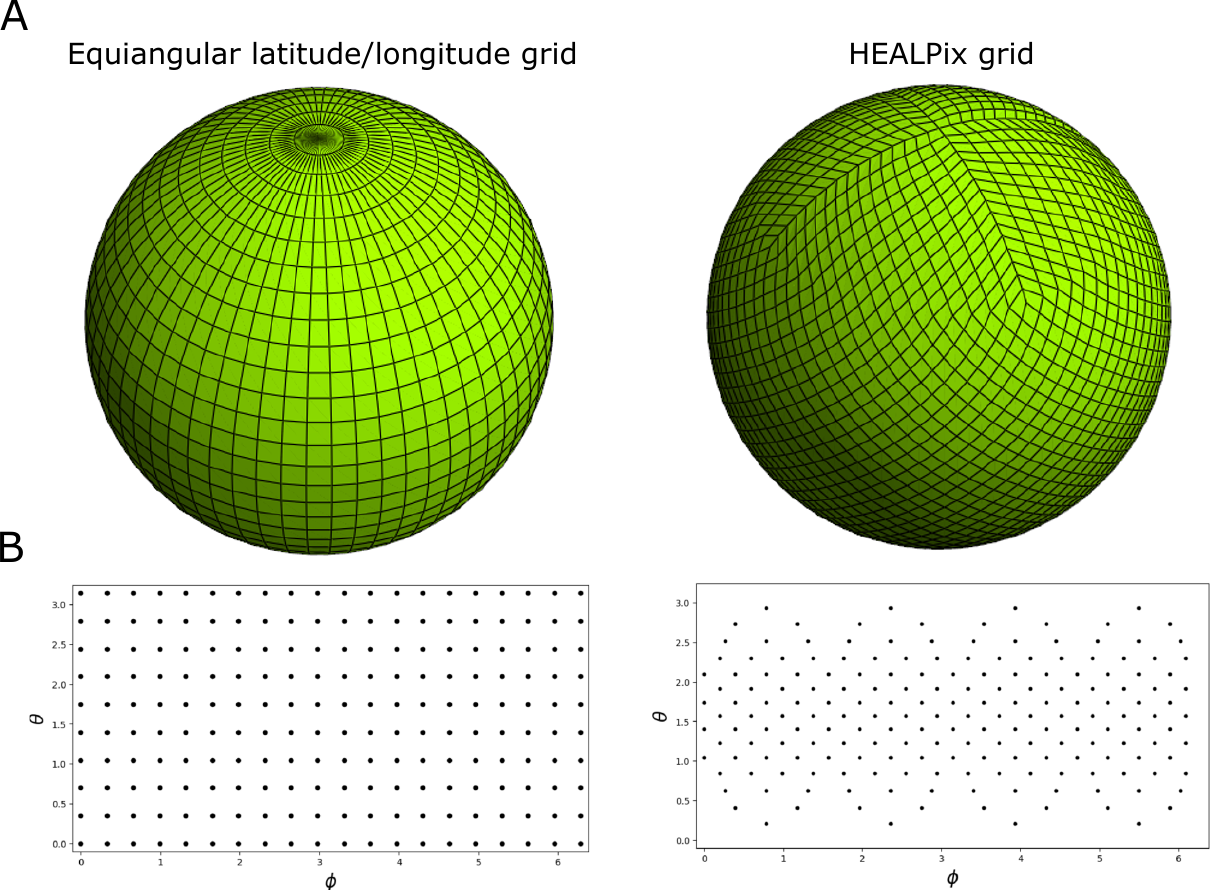}
    \caption{Visualization of (left) an equiangular latitude/longitude grid and (right) a HEALPix grid on (A) a sphere and (B) the latitude $\theta$ and longitude $\phi$ plane.}
    \label{fig:intro}
\end{figure}

To address these limitations, the Hierarchical Equal Area iso-Latitude Pixelization (HEALPix) scheme was introduced by Górski et al.~\cite{gorski2005healpix}. HEALPix divides the sphere into equal-area pixels while preserving an iso-latitude structure, as shown in Fig.~\ref{fig:intro}. Originally motivated by the need for precise analysis of CMB anisotropies, HEALPix provides a mathematically robust discretization of spherical functions that supports efficient and accurate statistical analysis of full-sky data. Today, HEALPix is widely adopted in astrophysics \cite{drake2020fast, reinecke2023improved} and Earth sciences \cite{karlbauer2024advancing}, where its balance of accuracy, scalability, and computational efficiency makes it well-suited for handling modern data-intensive applications.

Recently, the HEALPix grid has gained increasing attention in machine learning applications across climate science and astrophysics. 
Karlbauer et al.~\cite{karlbauer2024advancing} demonstrated that switching from cubed-sphere to HEALPix meshes enables parsimonious deep learning weather prediction models with global skill competitive to state-of-the-art architectures while using fewer prognostic variables. 
Krachmalnicoff and Tomasi~\cite{krachmalnicoff2019convolutional} introduced a pixel-based method for applying convolutional neural networks directly to HEALPix maps, showing that spherical CNNs can be integrated into mainstream ML frameworks and applied to both digit recognition and cosmological parameter inference from CMB data. Brenowitz et al.~\cite{brenowitz2025climate} recently advanced this direction with Climate in a Bottle (cBottle), a generative diffusion-based foundation model that emulates kilometer-scale global climate simulations directly on the HEALPix grid, achieving 5 km resolution and enabling zero-shot bias correction, downscaling, and data infilling while faithfully capturing extremes such as tropical cyclones and heatwaves. Mirowski et al.~\cite{mirowski2024neural} leveraged the area-preserving property of HEALPix to develop neural compression schemes for atmospheric states, achieving over 1000$\times$ compression ratios while faithfully retaining extreme events such as hurricanes and heatwaves. 
In astrophysics, Petroff et al.~\cite{petroff2020full} constructed a Bayesian spherical CNN on HEALPix maps to remove foreground contamination from full-sky millimeter observations, recovering the cosmic microwave background with high accuracy and per-pixel uncertainty estimates. 
More recently, Moreno et al.~\cite{moreno2024diagnosing} employed a U-Net architecture on HEALPix meshes to diagnose both typical and extreme precipitation events from reanalysis fields, highlighting the efficiency of HEALPix-based ML models for capturing localized extremes from coarse atmospheric data. 
Together, these studies illustrate that the iso-latitude and equal-area properties of HEALPix not only facilitate conventional statistical analyses but also provide a natural foundation for modern deep learning methods on spherical domains.

Despite these advances in discretization, the computational challenge of performing large-scale SHTs on HEALPix grids remains formidable. Existing implementations face scalability bottlenecks due to irregular data layouts, memory overheads, and limited parallelism. With emerging applications requiring the processing of trillions of pixels on massively parallel hardware, new methods are needed to fully exploit the potential of modern GPUs and distributed systems.

\textbf{Our contributions.} In this work, we introduce \texttt{cuHPX}, a GPU-accelerated framework for efficient and scalable SHTs and related utilities on HEALPix grids. Specifically, we make the following contributions:

\begin{itemize}

\item cuHPX library: We introduce \texttt{cuHPX}, a GPU-accelerated, differentiable library for spherical harmonic transforms and related utilities on HEALPix grids.

\item Kernel fusion strategy: We design a Bluestein-based batched FFT CUDA fused kernel tailored to HEALPix’s variable-length rings, improving efficiency on GPUs.

\item Differentiability support: We implement adjoint operators for analytical gradient computation, enabling seamless integration with PyTorch and other deep learning frameworks.

\item Out-of-core GPU transforms: We propose blocking strategies for Legendre transforms that reduce GPU memory usage while maintaining throughput.

\item Remapping utilities: We provide efficient routines for data reordering across RING, NESTED, and flat indexing layouts on HEALPix grids, facilitating interoperability with other spherical formats.

\item Performance results: On NVIDIA H100 GPUs, \texttt{cuHPX} achieves up to 20× speedup over \texttt{ducc} (20 CPU threads), 200× speedup over \texttt{healpy}, 7x speedup over GPU-based library \texttt{S2FFT}, while maintaining spectral convergence and numerical accuracy.

\end{itemize}

\section{Related Works}

Several libraries currently provide utilities for HEALPix, but most remain restricted to CPU-only execution. 
The most widely used package, \texttt{healpy}\cite{zonca2019healpy}, has become the standard tool in astrophysics for handling full-sky data and performing spherical harmonic analysis, but it offers no GPU acceleration. 
\texttt{ducc}\cite{reinecke2023improved} improves performance through multi-threaded CPU parallelism and supports general spherical harmonic transforms, yet it still cannot meet the demands of large-scale applications. 
More recently, \texttt{s2fft}\cite{price2024differentiable} introduced differentiable and accelerated spherical harmonic and Wigner transforms within the JAX ecosystem, enabling efficient gradient propagation and GPU scaling, although support for HEALPix remains limited. 
In parallel, NVIDIA’s \texttt{torch-harmonics} library~\cite{bonev2023spherical} implements differentiable spherical harmonics on equiangular latitude-longitude grids, which are suitable for deep learning applications but not directly applicable to HEALPix due to its misaligned longitude distribution (see Fig.~\ref{fig:intro}B). Wu et al.~\cite{wu2023anatomy,wu2025turbofft,wu2025turbofno} accelerate Fourier Neural Operator but lack support for Spherical Harmonic Transforms on HEALPix grids. Drake et al.~\cite{drake2020fast} proposed an algorithm for HEALPix-based transforms, demonstrating improved convergence rates and accuracy over traditional methods in CMB analysis; however, the implementation is not publicly available as open source.

These limitations highlight the need for GPU-accelerated methods that exploit massive parallelism while maintaining compatibility with modern deep learning frameworks. 
To this end, we developed \texttt{cuHPX}, a CUDA-optimized library that provides HEALPix utilities with several key features: 
(i) data reordering across different indexing layouts, 
(ii) optimized differentiable spherical harmonic transforms and their inverses on HEALPix grids, and 
(iii) efficient regridding between HEALPix and other spherical formats such as equiangular latitude-longitude grids. 
By bridging GPU acceleration with differentiable programming, \texttt{cuHPX} enables scalable applications in global climate modeling, astrophysics, and machine learning \cite{liu2024cusz,wu2025boosting}.

\subsection{Standard Methods in HEALPix Software \cite{reinecke2011libpsht}}

Implementing spherical harmonics on a HEALPix grid poses unique challenges compared to equiangular grids. 
In traditional CPU libraries, the spherical harmonic transform (SHT) and its inverse, known respectively as \emph{analysis} and \emph{synthesis}, are often formulated in matrix form, with operators $\mathbf{A}$ and $\mathbf{S}$. 
This allows the system to be solved iteratively or via gradient descent, with accuracy improved by increasing the number of iterations. 
However, a major drawback is that each iteration requires both an analysis and a synthesis step, leading to significant computational overhead.

Formally, for a continuous function $f$ with a spectral band limit of $\ell_{\max}$, the discrete analysis and synthesis are written as
\begin{equation}
a_{\ell}^m=\frac{4 \pi}{N} \sum_{i=1}^N \bar{Y}_{\ell}^m\left(\lambda_i, \theta_i\right) f\left(\lambda_i, \theta_i\right), \quad 
0 \leq \ell \leq \ell_{\max }, \, -\ell \leq m \leq \ell,
\end{equation}
\begin{equation}
\hat{f}\left(\lambda_i, \theta_i\right)=\sum_{\ell=0}^{\ell_{\max }} \sum_{m=-\ell}^{\ell} a_{\ell}^m Y_{\ell}^m\left(\lambda_i, \theta_i\right), 
\quad i=1, \ldots, N.
\end{equation}

where $a_{\ell}^m$ denotes spherical harmonic coefficients. 

In operator form, this becomes
\begin{equation}
\text{Analysis: } \mathbf{a} = \mathbf{A f}, \qquad
\text{Synthesis: } \hat{\mathbf{f}} = \mathbf{S a}.
\end{equation}

An iterative refinement can then be written as
\begin{equation}
\begin{aligned}
& \mathbf{r}^{(k+1)} = \mathbf{f} - \mathbf{S a}^{(k)}, \\
& \mathbf{a}^{(k+1)} = \mathbf{a}^{(k)} + \mathbf{A r}^{(k+1)},
\end{aligned}
\end{equation}
or equivalently, in gradient descent form:
\begin{equation}
\label{eq:iter-refine}
\mathbf{a}^{(k+1)} = \frac{4 \pi}{N} \mathbf{S}^* \mathbf{f} +
\left(\mathbf{I} - \frac{4 \pi}{N} \mathbf{S}^* \mathbf{S}\right) \mathbf{a}^{(k)}.
\end{equation}

The advantage of this iterative method is its controllable numerical accuracy. 
However, its computational cost scales as $O(\ell_{\max}^3)$, which quickly becomes prohibitive at high resolutions (e.g., $\ell_{\max} \gtrsim 2000$). 
This cubic scaling underscores the importance of GPU-accelerated approaches that leverage massive parallelism for tractable performance at modern problem sizes.

\section{Methods}

\subsection{Structure of HEALPix Grids on a Sphere}

One of the defining features of the HEALPix grid is its hierarchical subdivision and iso-latitude structure. These properties not only enable accurate spherical discretizations but also facilitate efficient parallel computations, making HEALPix particularly suitable for large-scale HPC and machine learning applications.

\begin{figure}[h]
    \centering
    \includegraphics[width=\linewidth]{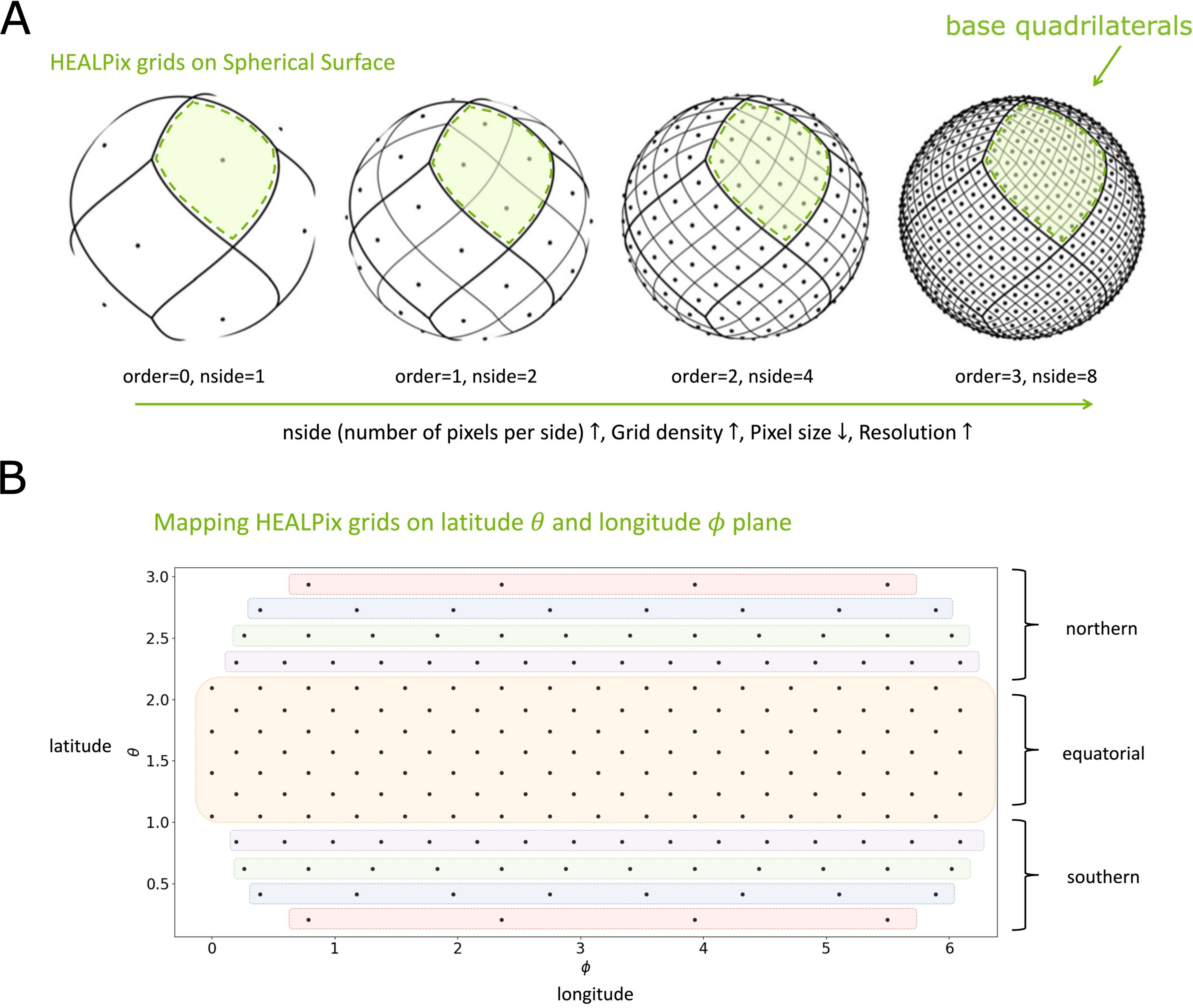}
    \caption{(A) Schematic diagram of hierarchical subdivision of HEALPix faces on a sphere. (B) Iso-latitude distribution of HEALPix Pixels on the latitude/longitude plane}
    \label{fig:intro2}
\end{figure}

The grid resolution is parameterized by \( n_{\text{side}} \), defined as a power of two. The total number of pixels on the sphere is given by $N_{\text{pix}} = 12 \times n_{\text{side}}^2$. The sphere is initially divided into 12 base quadrilaterals, each of which is recursively subdivided into four equal-area cells (Fig.~\ref{fig:intro2}A). As \( n_{\text{side}} \) increases, pixel sizes decrease accordingly, providing a higher resolution representation.

Another distinctive feature of HEALPix is its iso-latitude arrangement. Pixels are distributed along constant-latitude rings (Fig.~\ref{fig:intro2}B), resulting in a total of $N_{\text{rings}} = 4n_{\text{side}} - 1$
rings. Near the equator, each ring contains exactly \( 4n_{\text{side}} \) pixels, while near the poles the number of pixels per ring decreases, with only four pixels located at the rings closest to each pole. Although pixels within a ring are evenly spaced in longitude, their latitudinal positions are constant, creating an iso-latitude distribution. This property is unique to HEALPix and is essential for enabling FFT-based spherical harmonic transforms, which significantly accelerate spectral analysis on the sphere.

\subsection{Fast Data Reordering between Different Indexing Layouts}

\begin{figure*}[h]
    \centering
    \includegraphics[width=0.7\linewidth]{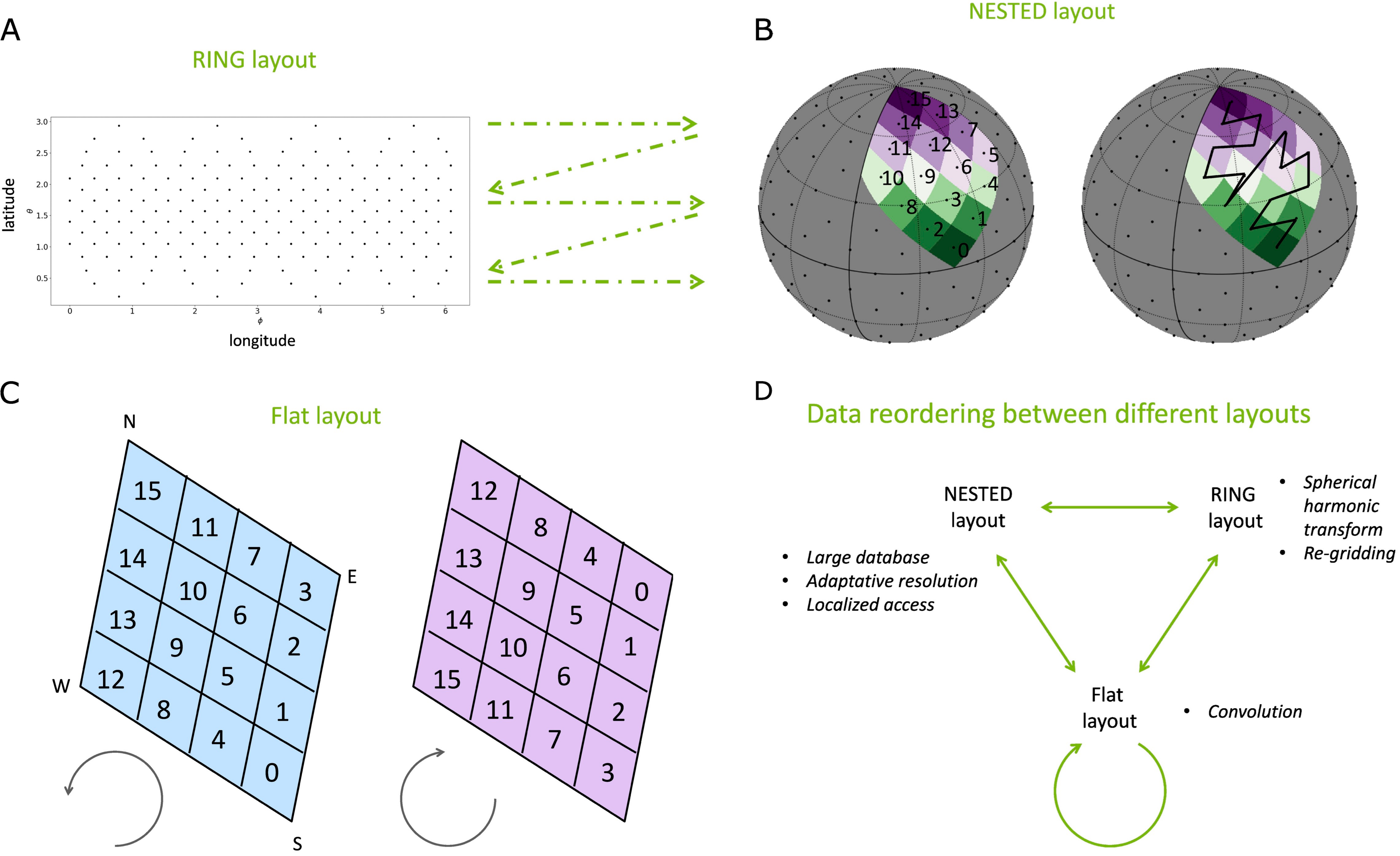}
    \caption{Schematic diagrams of (A) RING, (B) NESTED, (C) Flat indexing layout on HEALPix grids. Blue indicates counterclockwise ordering from the south, while pink indicates clockwise ordering from the east. (D) Data reordering on the HEALPix grids.}
    \label{fig:intro3}
\end{figure*}

Efficient indexing is critical for both data storage and large-scale computation on spherical grids. HEALPix supports multiple indexing layouts, each optimized for different applications:

\begin{itemize}

    \item \text{RING layout:} A format in which pixel indices increase from left to right along latitude rings, and rings are ordered from north to south (Fig.~\ref{fig:intro3}A). This layout is particularly convenient for spherical harmonic transforms, since FFT operations can be applied efficiently along the constant-latitude rings.

    \item \text{NESTED layout:} Exploits the hierarchical subdivision of HEALPix quadrilaterals. Pixel indices follow a Z-order curve (Morton order) within each base cell (Fig.~\ref{fig:intro3}B). This structure improves cache efficiency, supports adaptive resolution, and is well-suited for nearest-neighbor queries and database applications.

    \item \text{Flat layout:} Used in the convolution in HEALPix grid  (Fig.~\ref{fig:intro3}C), where indexing requires both an orientation (clockwise or counterclockwise) and a reference point. This layout simplifies the implementation of spherical convolutions \cite{karlbauer2024advancing} by leveraging existing convolution kernels for 2D images, particularly on GPUs.
\end{itemize}

These indexing formats are not only conventions for numbering pixels but also determine the order in which spherical data are stored and accessed. As shown in Fig.~\ref{fig:intro3}D, practical workflows often require reordering: for example, data generated in RING format via inverse spherical harmonic transforms may need to be converted into NESTED format for storage, or rearranged into flat layout for GPU convolution kernels. Thus, accurate and efficient reordering among layouts is crucial for enabling scalable parallel processing of spherical data.

Efficient conversion between HEALPix indexing schemes (e.g., RING, NESTED, and flat layouts) is essential for interoperability and high-performance computation. Different applications and kernels require data to be stored in specific layouts: spherical harmonic transforms typically use the RING format, hierarchical queries prefer the NESTED format, while convolution kernels on GPUs are simplified by the flat layout. To avoid writing pairwise converters between all possible formats, we introduce an intermediate representation, denoted as \texttt{xyf}. The \texttt{xyf} format encodes each pixel index as a triplet $(x, y, f)$, where $x$ and $y$ denote the two-dimensional coordinates within a base quadrilateral, and $f \in \{0, \dots, 11\}$ identifies the quadrilateral face on the sphere. Conversion between any two layouts can then be achieved in two steps:
\begin{enumerate}
    \item \text{Source $\rightarrow$ \texttt{xyf}:} Compute the $(x, y, f)$ tuple for each pixel based on the source indexing scheme.
    \item \text{\texttt{xyf} $\rightarrow$ Target:} Map each $(x, y, f)$ tuple into the desired target layout index.
\end{enumerate}
Finally, the data array is reordered according to the computed indices. This design reduces the complexity of supporting multiple layouts to maintaining only $O(n)$ conversion routines (to and from \texttt{xyf}), instead of $O(n^2)$ direct converters.

\begin{figure}[h]
    \centering
    \includegraphics[width=\linewidth]{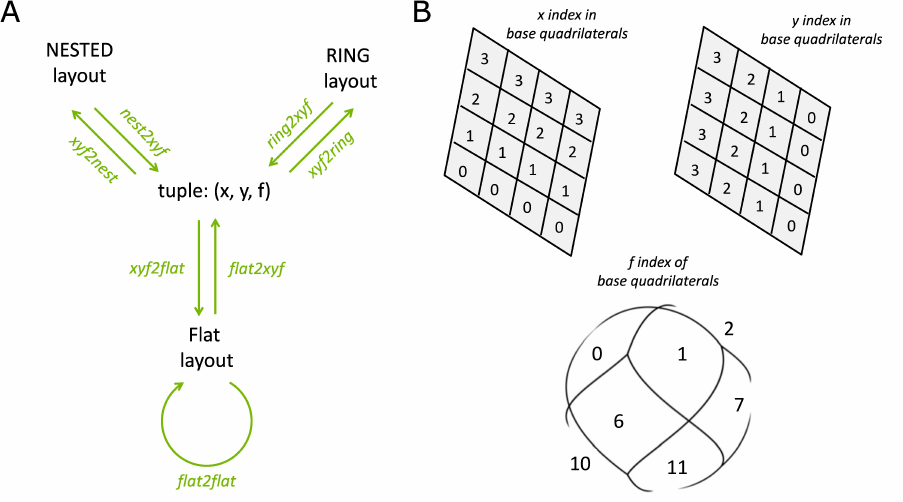}
    \caption{(A) Reordering paths between HEALPix indexing layouts via the \texttt{xyf} intermediate representation.
    (B) Illustration of the tuple $(x,y,f)$, where $x$ and $y$ denote local coordinates in a base quadrilateral, and $f$ identifies the face index.}
    \label{fig:method3}
\end{figure}

This intermediate representation greatly simplifies layout conversions and enables modular GPU implementations. Since each kernel only needs to understand how to convert between its preferred layout and \texttt{xyf}, the design naturally supports extensibility and minimizes redundant code paths.

\subsection{Overview of Spherical Harmonics}

Spherical harmonic transforms are fundamental tools for analyzing functions defined on the sphere and appear across many disciplines with spherical symmetry, including geophysics, astrophysics, quantum mechanics, computer graphics, and fluid dynamics \cite{cheng2023marginated, cheng2024microcirculatory, cheng2025red}. Their importance arises from the fact that spherical harmonics are eigenfunctions of the Laplace operator, which governs a wide range of physical systems.

Starting from Laplace’s equation
\begin{equation}
\nabla^2 \phi = 0,
\end{equation}
its expression in spherical coordinates yields the angular component
\begin{equation}
\frac{1}{\sin \theta} \frac{\partial}{\partial \theta}
\left( \sin \theta \frac{\partial Y}{\partial \theta} \right)
+ \frac{1}{\sin^2 \theta} \frac{\partial^2 Y}{\partial \phi^2}
+ \ell(\ell+1)Y = 0.
\end{equation}
Solving this equation by separation of variables proceeds via a Fourier
ansatz in longitude,
which reduces the $\theta$-dependence to a Sturm–Liouville eigenvalue problem
whose solutions are the associated Legendre polynomials $P_\ell^m(\cos\theta)$.
This yields the spherical harmonics,

\begin{equation}
Y_{\ell m}(\theta, \phi) =
\sqrt{\frac{2\ell+1}{4\pi}\frac{(\ell-m)!}{(\ell+m)!}}
\, P_\ell^m(\cos \theta)\, e^{i m \phi},
\end{equation}
where \(P_\ell^m(\cdot)\) are the associated Legendre polynomials. We denote the prefactor as
\(\lambda_\ell^m(\cos \theta)\) for notational convenience.

\begin{figure*}[h]
    \centering
    \includegraphics[width=0.7\linewidth]{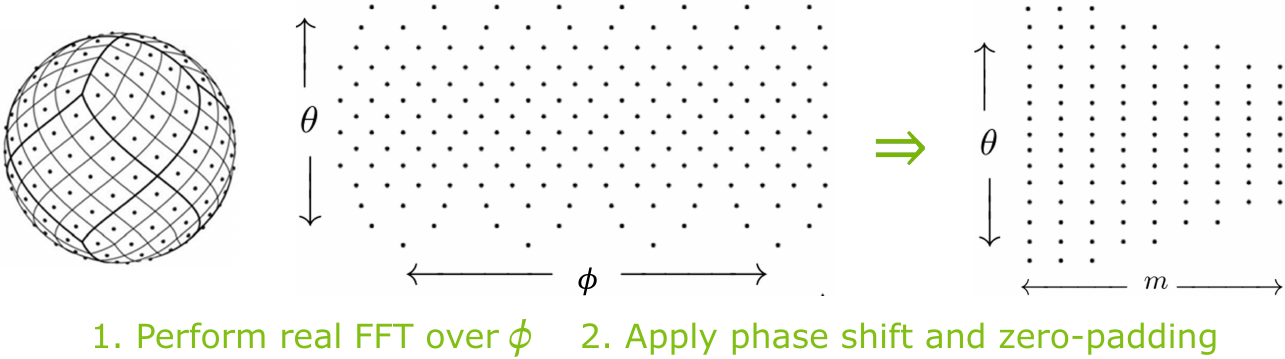}
    \caption{Efficient decomposition of spherical harmonic transforms on HEALPix.
    A real FFT (rFFT) is first performed in longitude (\(\phi\)) for each ring,
    followed by phase alignment, zero padding, and a Legendre transform in latitude (\(\theta\)).}
    \label{fig:method}
\end{figure*}

A critical property of spherical harmonics is their orthonormality over the sphere:
\begin{equation}
\int_{\mathbb{S}} Y_{\ell m}(\theta, \phi)
\, \overline{Y_{\ell^{\prime} m^{\prime}}(\theta, \phi)}\, d\Omega
= \delta_{\ell \ell^{\prime}} \delta_{m m^{\prime}}.
\end{equation}
This property allows any square-integrable function \( f(\theta,\phi) \) on the sphere to be expressed as a linear combination of spherical harmonics.

The \textit{forward} spherical harmonic transform (analysis) projects \( f \) onto the basis:
\begin{equation}
a_{\ell m} = \int_{\mathbb{S}} f(\theta, \phi)\,
\overline{Y_{\ell m}(\theta, \phi)} \, d\Omega,
\end{equation}
where \( a_{\ell m} \) are the spectral coefficients. Conversely, the
\textit{inverse} transform (synthesis) reconstructs \( f \) from these coefficients:
\begin{equation}
f(\theta, \phi) =
\sum_{\ell=0}^{\ell_{\max}} \sum_{m=-\ell}^{\ell}
a_{\ell m} Y_{\ell m}(\theta, \phi).
\end{equation}

The challenge in implementing spherical harmonic transforms efficiently lies in the discretization of the sphere. On HEALPix grids, while pixels are iso-latitude, they are unevenly distributed along lines of constant longitude (Fig.~\ref{fig:intro}B). This misalignment complicates direct FFT application in the longitudinal direction, necessitating specialized algorithms and data layouts to achieve scalability on high-performance architectures.

\subsection{Efficient Spherical Harmonic Transform}

Direct evaluation of spherical harmonic transforms is computationally expensive, as it involves a two-dimensional integral over the sphere. Instead, more efficient formulations decompose the transform into a sequence of one-dimensional operations \cite{schaeffer2013efficient}. Specifically, the computation can be separated into: (1) a Fourier transform in longitude (\(\phi\)), followed by (2) a Legendre transform in latitude (\(\theta\)).

The spherical harmonic coefficients are formally defined as
\begin{equation}
f_n^m = \int_0^{2 \pi} \int_0^\pi f(\theta, \phi)\,
\overline{Y_n^m}(\theta, \phi)\, \sin \theta \, d\theta \, d\phi.
\end{equation}
By first applying a Fourier transform in \(\phi\), we obtain
\begin{equation}
f_m(\theta) = \int_0^{2 \pi} f(\theta, \phi) e^{-im\phi} d\phi,
\end{equation}
which reduces the problem to a Legendre transform along \(\theta\):
\begin{equation}
f_n^m = \sum_{j=1}^{N_\theta} f_m(\theta_j)\,
P_n^m(\cos \theta_j)\, w_j,
\end{equation}
where \(w_j\) are quadrature weights. An illustration of this decomposition is shown in Fig.~\ref{fig:method}.

This formulation provides several computational advantages for high-performance implementations:
\begin{itemize}
    \item \text{FFT acceleration:} The Fourier step can be computed using standard FFT libraries such as \texttt{cuFFT}, enabling efficient GPU parallelization.
    \item \text{Hermitian symmetry:} Since the input data are real-valued, Hermitian symmetry reduces both compute and memory requirements by nearly a factor of two.
    \item \text{Batch processing:} Thousands of FFTs (of varying lengths) are required, which can be grouped into batched executions to improve throughput on GPUs.
    \item \text{Zero padding:} Applying zero padding in the Fourier domain ensures consistency across FFT lengths and smooths spectral leakage effects.
    \item \text{Precomputation:} Legendre polynomials \(P_n^m(\cos \theta)\) can be precomputed and stored, amortizing cost across multiple transforms.
\end{itemize}

The Legendre functions are defined as
\begin{equation}
P_n^m(x) =
(-1)^m \sqrt{\frac{2n+1}{4\pi}}
\sqrt{\frac{(n-|m|)!}{(n+|m|)!}}
(1-x^2)^{|m|/2} \frac{d^{|m|}}{dx^{|m|}} P_n(x),
\end{equation}
and can be generated efficiently using recurrence relations:
\begin{equation}
P_n^m(x) = a_n^m x P_{n-1}^m(x) + b_n^m P_{n-2}^m(x),
\end{equation}
where coefficients \(a_n^m\) and \(b_n^m\) depend only on \((n,m)\).

In summary, this two-step formulation, Fourier transform in longitude followed by a Legendre transform in latitude, is well-suited for modern HPC systems. Its reliance on FFT libraries, recursive polynomial evaluation, and batched parallelism makes it particularly effective for GPU acceleration on HEALPix grids.

\subsection{Bluestein's FFT for Kernel Fusion on GPUs}

The Bluestein FFT \cite{swarztrauber1991bluestein, bluestein2003linear}, also known as the chirp-$z$ transform, extends the fast Fourier transform to arbitrary input sizes, not restricted to powers of two. This property makes it particularly suitable for HEALPix grids, where the number of points per latitude ring generally varies and is often non-power-of-two. Bluestein’s algorithm reformulates the discrete Fourier transform (DFT) as a convolution, which can be evaluated efficiently using FFTs.

The DFT of a sequence $x[n]$ of length $N$ is
\begin{equation}
X(k) = \sum_{n=0}^{N-1} x[n] W_N^{nk}, \qquad
W_N = e^{-2\pi i / N}.
\end{equation}
This can be rewritten as
\begin{equation}
X(k) = \sum_{n=0}^{N-1} x[n] W_N^{n^2/2} W_N^{-(k-n)^2/2} W_N^{k^2/2}.
\end{equation}
Defining the chirp sequence
\begin{equation}
h[m] = W_N^{m^2/2},
\end{equation}
the DFT is equivalently expressed as
\begin{equation}
X(k) = h[k] \cdot \Bigg( \sum_{n=0}^{N-1} \big( x[n] h[n] \big) \, h^{-1}[k-n] \Bigg).
\end{equation}

This formulation reduces the DFT to three main steps, each of which can be mapped efficiently to GPU kernels:
\begin{enumerate}
    \item \text{Chirp multiplication:}
    \begin{equation}
    y[n] = x[n] \, h[n].
    \end{equation}
    \item \text{Convolution with inverse chirp:}
    \begin{equation}
    w[k] = \sum_{n=0}^{N-1} y[n] \, h^{-1}[k-n],
    \end{equation}
    implemented via FFT $\rightarrow$ pointwise multiply $\rightarrow$ inverse FFT.
    \item \text{Final chirp multiplication:}
    \begin{equation}
    X(k) = w[k] \, h[k].
    \end{equation}
\end{enumerate}

\begin{figure}[h]
    \centering
    \includegraphics[width=\linewidth]{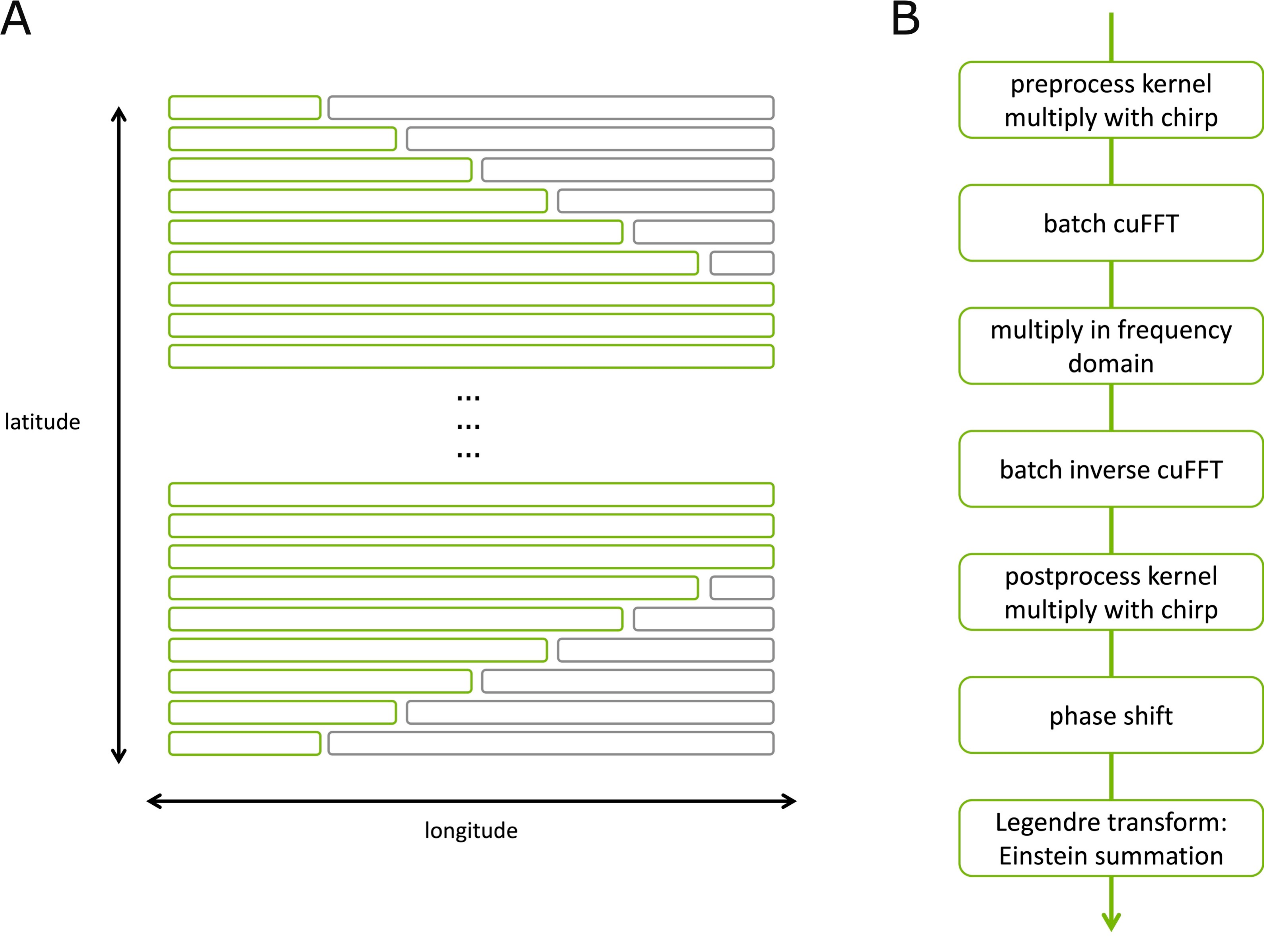}
    \caption{(A) Illustration of Bluestein's FFT on HEALPix rings. Green elements denote data samples, and gray padding ensures a consistent chirp length across rings. 
    (B) Flow of the spherical harmonic transform based on Bluestein’s FFT, enabling efficient batched execution on GPUs.}
    \label{fig:method2}
\end{figure}

This approach preserves the $O(N \log N)$ complexity of the FFT while allowing transforms of arbitrary lengths. More importantly, using chirps of uniform length across HEALPix rings enables all convolutions (step~2) to be executed as a single batched \texttt{cuFFT} operation. This strategy fuses multiple transforms into a unified kernel, significantly reducing kernel launch overhead and memory traffic. As a result, Bluestein’s FFT provides both flexibility and high throughput, making it well-suited for scalable GPU acceleration of spherical harmonic transforms on HEALPix grids.

\subsection{Adjoint Methods for Differentiability}

The adjoint state method is a powerful technique for efficiently computing gradients in optimization problems governed by linear or nonlinear operators \cite{plessix2006review, dwight2006effect}. It is widely applied in physics, control theory, photonics, and more recently in machine learning, e.g., for neural ODEs \cite{chen2018neural, linot2023stabilized}. The key idea is to reformulate gradient computation as a constrained optimization problem, and then derive its dual form. This avoids explicitly forming Jacobians and yields gradients at a cost comparable to a single forward evaluation.

\paragraph{SHT and iSHT as Linear Operators \cite{reinecke2013libsharp}} 
The synthesis (inverse spherical harmonic transform, iSHT) can be expressed as 
\begin{equation}
\boldsymbol{p}_{xy} = 
\sum_{m=-\ell_{\max}}^{\ell_{\max}}
\sum_{\ell = |m|}^{\ell_{\max}}
a_{\ell m}\,
\lambda_{\ell m}(\vartheta_y)\,
\exp\!\left(i m \varphi_{0,y} + \frac{2\pi i m x}{N_{\varphi,y}}\right),
\end{equation}
which can be written compactly in operator form as
\begin{equation}
\boldsymbol{p} = Y \boldsymbol{a}.
\end{equation}
Here, \(Y\) is the iSHT operator mapping spectral coefficients \(\boldsymbol{a}\) to spatial values \(\boldsymbol{p}\).

Similarly, the analysis (forward spherical harmonic transform, SHT) can be written as
\begin{equation}
\hat{a}_{\ell m} =
\sum_{y=0}^{N_s-1} \sum_{x=0}^{N_{\varphi,y}-1}
\boldsymbol{p}_{xy}\, w_{ys}\, \lambda_{\ell m}(\vartheta_y)\,
\exp\!\left(-i m \varphi_{0,y} - \frac{2\pi i m x}{N_{\varphi,y}}\right),
\end{equation}
or equivalently,
\begin{equation}
\boldsymbol{a} = Y^{\dagger} W \boldsymbol{p},
\end{equation}
where \(W\) is a diagonal matrix of quadrature weights, assumed to be invertible.

\paragraph{Adjoint Operators} 
Since the SHT and iSHT are linear, their adjoints are well defined and play a central role in gradient-based optimization. In particular:
\begin{itemize}
    \item The adjoint iSHT is given by \( Y^{\dagger} \), which maps spatial fields back to spectral space.
    \item The adjoint SHT is given by \( W Y \), which applies quadrature weights before the forward projection.
\end{itemize}
Importantly, the adjoint is not the same as the inverse, but is instead the operator required to compute sensitivities in optimization frameworks. For example, when optimizing a cost function defined in the spatial domain with respect to spectral coefficients, the adjoint iSHT provides an efficient way to propagate gradients.

\paragraph{Implementation.}
In practice, one can obtain \(Y^{\dagger}\) by leveraging an existing iSHT implementation, \(Y^{\dagger} W\), and multiplying by \(W^{-1}\) \cite{reinecke2013libsharp}. This reuse of forward and inverse transforms makes adjoint computations nearly cost-free compared to the original transform, enabling scalable gradient-based optimization on spherical domains.

\begin{figure*}[h]
    \centering
    \includegraphics[width=\linewidth]{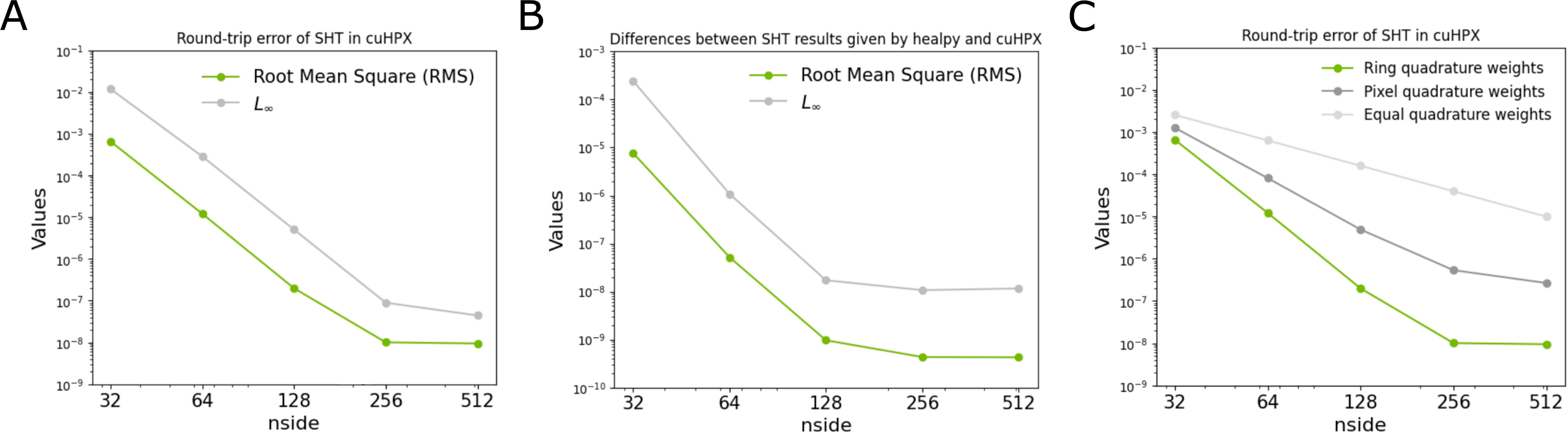}
    \caption{(A) Round-trip error of spherical harmonic transform in \texttt{cuHPX}, showing spectral convergence. 
    (B) Agreement between spherical harmonic transform results from \texttt{healpy} and \texttt{cuHPX}, converging to machine precision. 
    (C) Round-trip error for different choices of quadrature weights, where ring weights yield superior accuracy.}
    \label{fig:result_accuracy}
\end{figure*}

\subsection{Out-of-core Memory Optimization for Legendre Transform}

The Legendre transform step of the spherical harmonic transform is defined as
\begin{equation}
f_n^m = \sum_{j=1}^{N_\theta} f_m(\theta_j)\, P_n^m(\cos \theta_j)\, w_j,
\end{equation}
where $N_\theta$ denotes the number of latitude rings. Unlike the Fourier step, which can be accelerated with standard FFT libraries, the main challenge here lies in the memory footprint of the Legendre weight tensor $P$. The tensor scales as
$\mathcal{O}(N_\theta \, \ell_{\max} \, m_{\max}),$ which for typical resolutions (e.g., $n_{\text{side}}=1024$, $\ell_{\max}=2048$) exceeds 70~GB. This likely surpasses the available memory of GPUs, making efficient memory management essential.

To overcome this bottleneck, we adopt a streaming strategy inspired by the \texttt{einsum} formulation of the Legendre transform. The weight tensor is partitioned along the one dimension into multiple chunks. These chunks are processed sequentially in a pipeline overlapping compute and memory transfers using CUDA streams as follows:
\begin{itemize}
    \item While chunk $i$ is being transferred from pinned host memory to the GPU, computation is performed on chunk $i-1$.
    \item Similarly, when chunk $i+1$ is queued for transfer, chunk $i$ is simultaneously processed on the GPU.
\end{itemize}
This overlap of data transfer and computation ensures that GPU resources are continuously utilized while keeping the peak device memory footprint bounded by the chunk size rather than the full tensor.

This chunked streaming approach achieves two critical optimizations:
\begin{enumerate}
    \item {Reduced memory footprint:} Only a fraction of the tensor resides in GPU memory at any given time, enabling transforms at scales otherwise infeasible.
    \item {Improved throughput:} Overlapping host-device transfers with computation hides memory latency and ensures near-saturated GPU utilization.
\end{enumerate}
As a result, the proposed method enables practical large-scale spherical harmonic transforms on high-resolution HEALPix grids, balancing memory efficiency and computational performance.

\section{Results \& Discussion}

\subsection{Accuracy \& Convergence}

To ensure the correctness of our implementation, we first evaluate the \emph{round-trip error}, defined as
\begin{equation}
\| f - i\!SHT(SHT(f)) \|, 
\end{equation}
where both the root-mean-square (RMS, or $L_2$) norm and the maximum ($L_\infty$) norm are reported. Here, $f$ is taken to be band-limited to 
$\ell_{\max} = 3n_{\text{side}} - 1$, which is the bandwidth used by SHT on the 
HEALPix grids. As shown in Fig.~\ref{fig:result_accuracy}(A), the error decreases rapidly with increasing $nside$, demonstrating the expected spectral convergence of the algorithm. 
At $nside = 256$, the error is already reduced to machine precision, indicating that the implementation does not introduce numerical artifacts.

Fig.~\ref{fig:result_accuracy}(B) further compares the results produced by \texttt{cuHPX} and the widely used \texttt{healpy} package. 
The difference between the two implementations quickly vanishes to machine precision as resolution increases, confirming that the GPU-based kernels in \texttt{cuHPX} faithfully reproduce the mathematical transform. 
This result is particularly important since it establishes numerical consistency with the standard CPU-based library while offering significant performance improvements.

Beyond correctness, it is critical to understand how discretization choices influence convergence. 
In particular, the choice of quadrature weights directly affects the accuracy of the Legendre transform on the HEALPix grid \cite{swarztrauber2003computing}. 
The Legendre transform in continuous form is
\begin{equation}
f_n^m = \int_0^\pi f_m(\theta) P_n^m(\cos \theta) \sin \theta \, d\theta,
\label{eq:legendre_cont}
\end{equation}
which in discrete form becomes
\begin{equation}
f_n^m = \sum_{j=1}^{N_\theta} f_m(\theta_j) P_n^m(\cos \theta_j) w_j,
\label{eq:legendre_disc}
\end{equation}
where $w_j$ are the quadrature weights. 

We evaluate three commonly used weighting schemes: 
\text{equal} (all pixels share the same weight), 
\text{ring} (all pixels in a ring share the same weight), 
and \text{pixel} (each pixel has a distinct quadrature weight). 
As shown in Fig.~\ref{fig:result_accuracy}(C), the choice of quadrature weights has a direct and measurable impact on convergence. 
Equal weights lead to visible stagnation in accuracy, while pixel weights provide moderate improvement. 
Ring weights, however, provide the best results, reducing the round-trip error by nearly an order of magnitude compared to equal weighting at the same resolution. 
This analysis highlights that accuracy on the HEALPix grid is not only a function of grid resolution $nside$ but also of careful algorithmic design choices. 
Consequently, \texttt{cuHPX} adopts ring weights as the default, striking a balance between computational simplicity and spectral accuracy.

\subsection{Performance Benchmark \& Comparison}

\subsubsection{Data Remapping}

\begin{figure}[h]
    \centering
    \includegraphics[width=\linewidth]{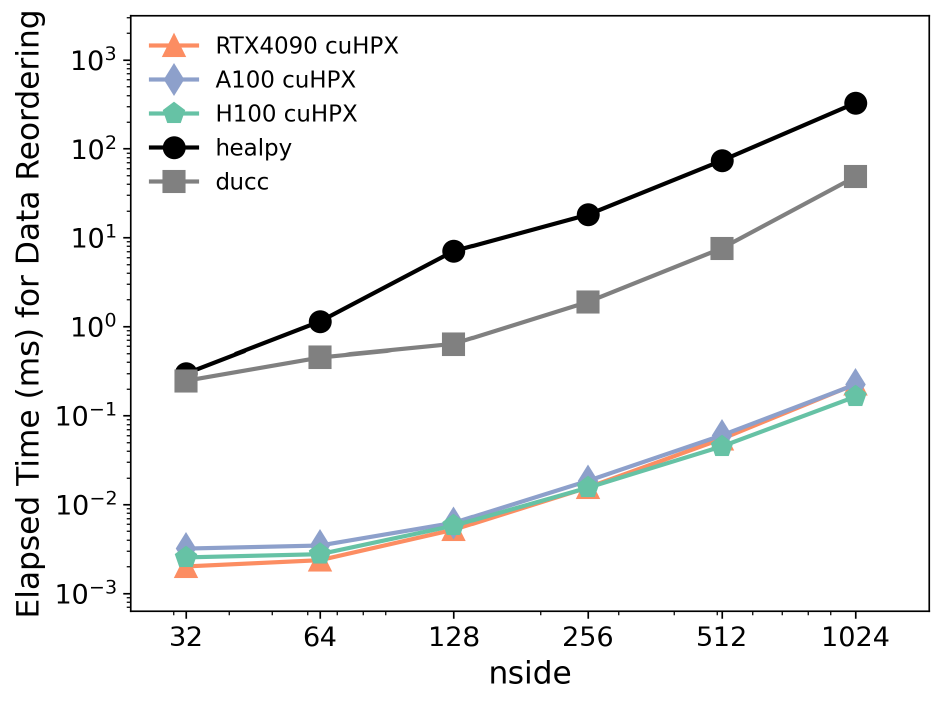}
    \caption{Elapsed Time (milliseconds) used by data reordering from NESTED layout to RING layout on HEALPix grids. \texttt{healpy} uses a single CPU thread, \texttt{ducc} uses 20 CPU threads in parallel on AMD EPYC 9454.}
    \label{fig:result_time_remap}
\end{figure}

We also compare the performance on data reordering of \texttt{cuHPX} with \texttt{healpy} and \texttt{ducc}, analyzing both batched and non-batched scenarios with timing data summarized in Fig.~\ref{fig:result_time_remap}. Unlike \texttt{healpy} and \texttt{ducc}, which lack native support for batch processing and rely on external loops for non-batched execution on CPU, \texttt{cuHPX} supports batch operations directly. Benchmarks for \texttt{cuHPX} were performed on NVIDIA RTX 4090, A100, and H100 GPUs, where it achieved the highest speedups. For $n_{\text{side}} = 1024$, \texttt{cuHPX} demonstrated a 300x speedup on the H100 GPU compared to the multi-threaded \texttt{ducc} and a 2000x speedup relative to the single-threaded \texttt{healpy}, showcasing its efficiency for computationally intensive tasks.

\begin{figure}[h]
    \centering
    \includegraphics[width=\linewidth]{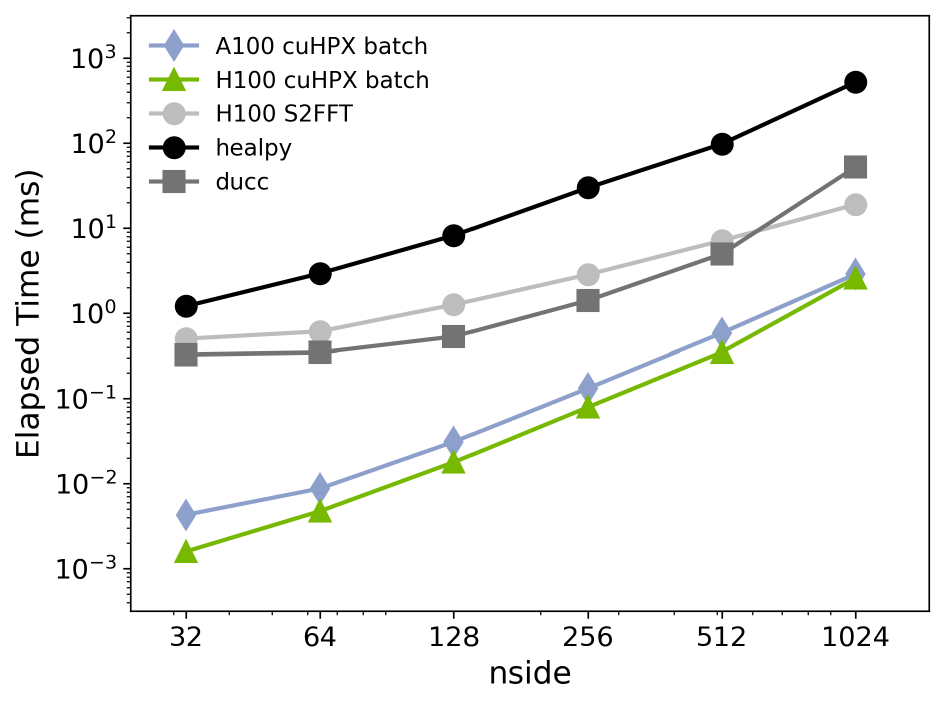}
    \caption{Elapsed Time (milliseconds) used by Spherical Harmonics Transforms on HEALPix grids. \texttt{healpy} uses a single CPU thread, \texttt{ducc} uses 20 CPU threads in parallel on AMD EPYC 9454.}
    \label{fig:result_time_sht}
\end{figure}

\subsubsection{Spherical Harmonic Transforms}

We compared the performance of spherical harmonic transforms across \texttt{cuHPX}, \texttt{S2FFT}, \texttt{healpy}, and \texttt{ducc}. The \texttt{healpy} library was evaluated in a single-threaded CPU configuration, while \texttt{ducc} utilized 20 CPU threads in parallel. \texttt{S2FFT} is a Python package for computing differentiable Fourier transforms on the sphere, designed for deployment on hardware accelerators \cite{price2024differentiable}. For \texttt{cuHPX}, benchmarks were performed on both NVIDIA A100 and H100 GPUs. At \(n_{\text{side}} = 1024\), \texttt{cuHPX} achieved a speedup of approximately $20\times$ on the H100 GPU compared to the multi-threaded \texttt{ducc} implementation and nearly $200\times$ relative to the single-threaded \texttt{healpy}, as summarized in Fig~\ref{fig:result_time_sht}. Moreover, \texttt{cuHPX} provided a $7\times$ performance improvement over \texttt{S2FFT} on the same H100 hardware.

\subsection{Differentiability}

In addition to raw performance, differentiability is a key requirement for modern scientific machine learning applications. 
Differentiability ensures that \texttt{cuHPX} can provide gradients of spherical harmonic transforms for backpropagation, thereby enabling seamless integration with deep learning frameworks. 
This capability is particularly relevant for tasks such as spherical convolutional neural networks, physics-informed neural operators, and differentiable rendering, where gradients with respect to spherical functions must be computed efficiently.

To validate differentiability, we constructed a simple neural network consisting of a single trainable layer. 
The loss function was defined as 
\[
\mathcal{L} = \big(y - i\!SHT(x)\big)^2,
\]
where $y$ is a target signal on the sphere and $x$ represents trainable spherical harmonic coefficients initialized with random values. 
During training, \texttt{cuHPX} provides analytic gradients: the gradient of the spherical harmonic transform is computed via the adjoint inverse spherical harmonic transform, while the gradient of the inverse transform is computed via the adjoint spherical harmonic transform, as derived in the previous section. 
This approach avoids the overhead of generic automatic differentiation frameworks by exploiting operator adjointness directly.

\begin{figure*}[h]
    \centering
    \includegraphics[width=\linewidth]{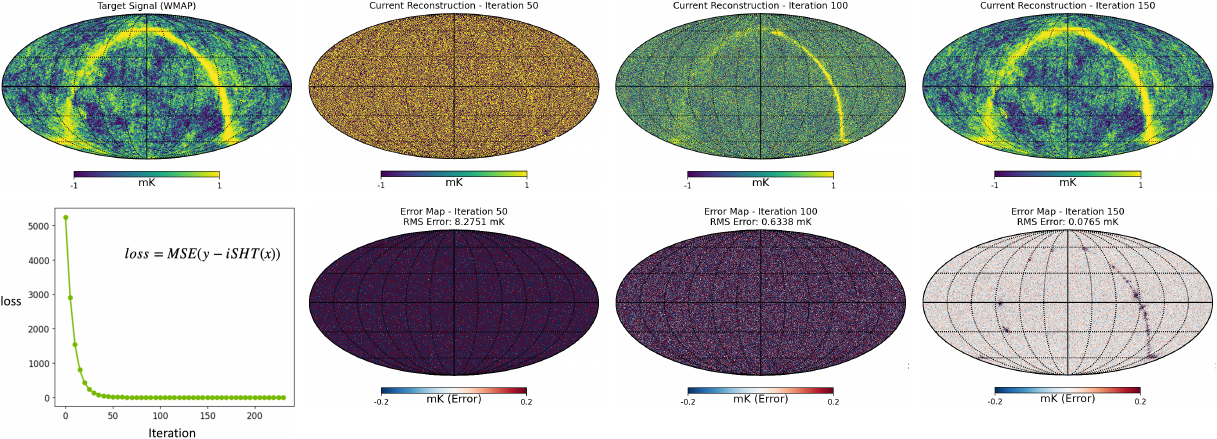}
    \caption{Loss curve during gradient descent training using \texttt{cuHPX}. 
    The loss is defined as $\mathcal{L} = MSE(y - i\!SHT(x))$, where $y$ is the target spherical signal, $x$ is initialized as a random complex variable, and $MSE$ denotes mean square error loss. 
    Gradients are computed analytically using adjoint operators rather than generic automatic differentiation. 
    The rapid convergence of the model validates both the correctness and stability of the gradient implementation. The target signal is a scientific sky map from Wilkinson Microwave Anisotropy Probe (WMAP) 7-year data release \cite{jarosik2011seven}, containing the microwave intensity and polarization of the cosmic microwave background in the W band.}
    \label{fig:result_diff}
\end{figure*}

Fig.~\ref{fig:result_diff} shows the training loss curve obtained using gradient descent. 
The rapid and monotonic decrease of the loss function indicates stable convergence, which in turn verifies the correctness of the analytic gradients implemented in \texttt{cuHPX}. 
Equally important, the smooth loss trajectory confirms that the adjoint operators are numerically well-conditioned, ensuring stable gradient propagation even for high-resolution inputs.
These results demonstrate that \texttt{cuHPX} is not only accurate and high-performing but also differentiable. 
By providing efficient analytic gradients, \texttt{cuHPX} extends its utility beyond numerical simulation into emerging applications that couple spherical harmonics with machine learning, including differentiable physics solvers and neural operators.  This positions \texttt{cuHPX} as a bridge between high-performance scientific computing and gradient-based machine learning frameworks.

\begin{figure*}[h]
    \centering
    \includegraphics[width=0.9\linewidth]{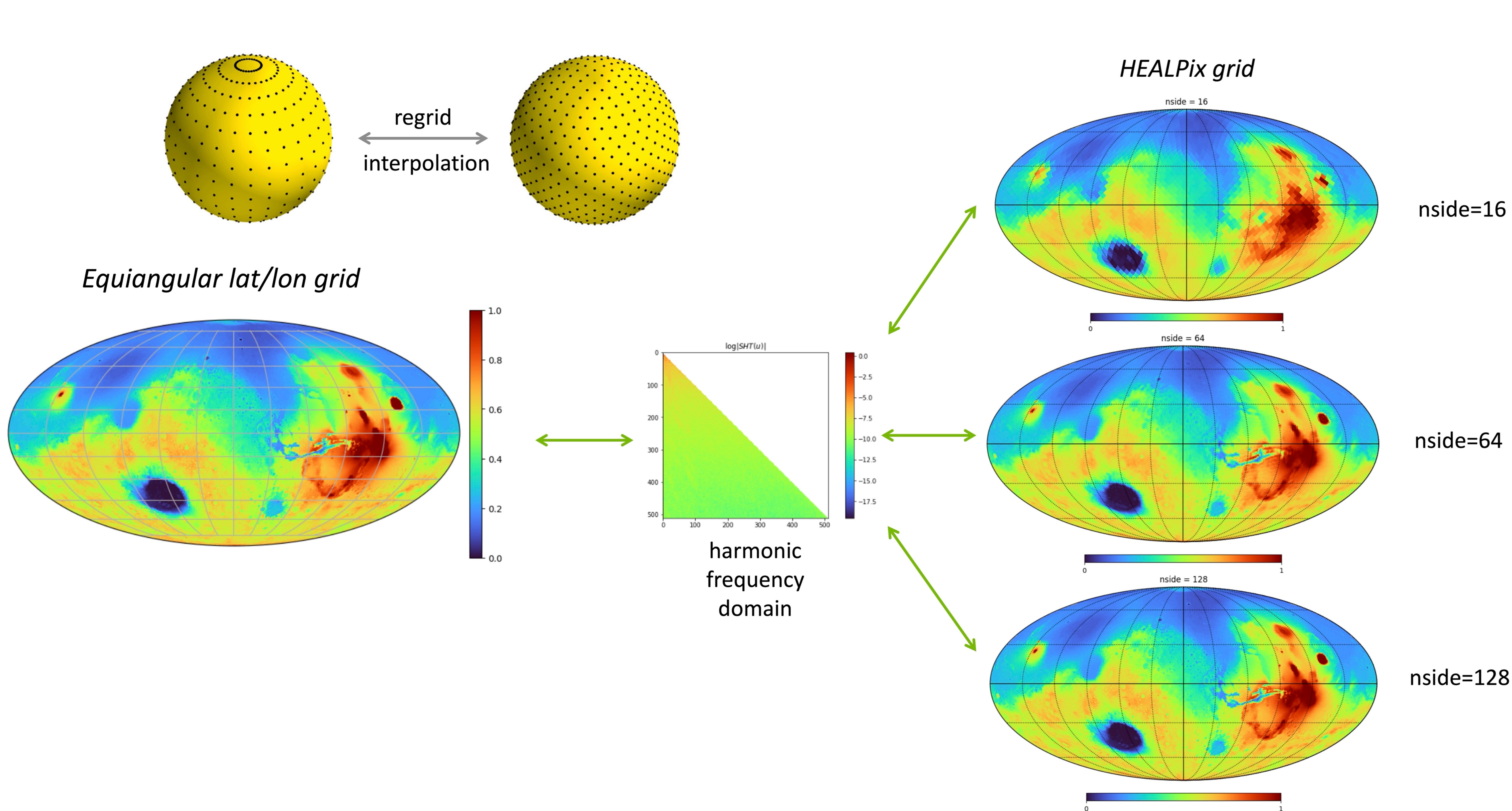}
    \caption{Schematic diagram of regridding between equiangular latitude-longitude and HEALPix, as well as upsampling and downsampling via spherical harmonic transform and its inverse. The original plot is a color-coded elevation map of Mars, derived from laser altimetry, showing the planet’s major mountains, basins, and global dichotomy between northern lowlands and southern highlands \cite{smith2001mars}.}
    \label{fig:result_regrid}
\end{figure*}

\subsection{Regridding}

A common challenge in large-scale geophysical and astrophysical simulations is the need to exchange data between different spherical grid representations. 
For example, global climate models often employ equiangular latitude--longitude grids, while astrophysical analyses frequently rely on HEALPix for equal-area pixelization. 
Direct interpolation between these grids can introduce non-trivial aliasing errors and is computationally expensive at high resolutions. 
\texttt{cuHPX} addresses this challenge by enabling regridding through spherical harmonic transforms, which ensures both accuracy and efficiency. 

Specifically, data sampled on an equiangular latitude--longitude grid can first be transformed into spectral space using \texttt{torch-harmonics}~\cite{bonev2023spherical}. 
The spectral coefficients are then mapped back to the spatial domain on the HEALPix grid using the inverse spherical harmonic transform (iSHT) provided by \texttt{cuHPX}. 
This pipeline avoids pointwise interpolation and instead relies on mathematical basis function expansions, guaranteeing spectral accuracy provided the input resolution is sufficiently high. 
The same mechanism also allows upsampling and downsampling of HEALPix data by truncating or extending spherical harmonic coefficients before applying the iSHT, as illustrated in Fig.~\ref{fig:result_regrid}. 

Beyond equiangular and HEALPix grids, the same approach generalizes to other quadrature-based spherical discretizations such as Legendre–Gauss or Lobatto grids. 
Thus, the SHT and iSHT operators in \texttt{cuHPX} serve as a unifying bridge among heterogeneous spherical grid formats. 
This capability is particularly valuable in hybrid workflows where simulation, analysis, and machine learning components adopt different discretizations. 

In summary, regridding via \texttt{cuHPX} provides three main advantages:  
(1) it preserves spectral accuracy compared to ad-hoc interpolation methods;  
(2) it scales efficiently on GPUs, making high-resolution data exchange feasible in practice; and  
(3) it enables seamless coupling between different numerical solvers and data-driven models operating on distinct grid structures. 
These features position \texttt{cuHPX} as a versatile tool for integrating diverse computational frameworks in large-scale spherical applications.

\section{Limitation \& Future Work}

One important direction is to extend \texttt{cuHPX} to support multi-GPU and distributed spherical harmonic transforms. 
At high resolutions ($nside \geq 4096$), both memory and computational demands exceed the capacity of a single GPU. 
A distributed SHT requires partitioning the input data across devices, overlapping communication and computation, and exploiting high-bandwidth GPU interconnects (e.g., NVLink, or Infiniband) to efficiently exchange intermediate spectral coefficients. 
A natural strategy is to distribute subsets of rings or $m$-modes across GPUs, allowing local FFTs and Legendre transforms to be performed independently before performing an all-to-all communication step to assemble the global spectral result. 
This multi-GPU extension would not only enable scaling to ultra-high-resolution HEALPix grids, but also integrate seamlessly with large-scale HPC clusters, making \texttt{cuHPX} suitable for production cosmology pipelines and global climate simulations.

In addition, a promising avenue is to leverage \texttt{cuHPX} as the computational core for \text{solving physical PDEs on the sphere}, such as the shallow water equations on HEALPix grids. 
Shallow water dynamics are a canonical benchmark for geophysical fluid dynamics, capturing key features of large-scale atmospheric and oceanic flows. 
By integrating \texttt{cuHPX}’s efficient spherical harmonic transforms with time-stepping schemes, one could implement a fully GPU-accelerated spectral solver that evolves prognostic variables on the HEALPix grid. 
Such a solver would demonstrate both the numerical accuracy of \texttt{cuHPX} and its potential for real-world scientific applications, bridging the gap between low-level GPU kernels and end-to-end models used in climate science and planetary simulations.

Another important extension is to explore the accuracy performance trade-offs through iterative refinement methods, such as the approach described in Eq.~\ref{eq:iter-refine}. Because HEALPix grids do not support exact spherical harmonic transforms, small aliasing and discretization errors accumulate in practice. Iterative refinement offers a principled way to reduce these errors, allowing users to dial in the level of accuracy needed for a given application while controlling the additional computational cost. This flexibility could be particularly valuable in workflows where extreme precision is essential, such as cosmological inference or the diagnosis of rare extreme events in climate simulations.

\section{Conclusion}

The development and optimization of SHT on HEALPix grids represent advancements in computational efficiency and applicability across fields requiring spherical data processing. The \texttt{cuHPX} library advances differentiable SHT on HEALPix grids and facilitates adaptive regridding across different grid types, leveraging GPU-accelerated computation for enhanced performance and enabling seamless integration with deep learning frameworks. Key optimizations include kernel fusion for batched FFTs based on the Bluestein algorithm, strategies to address HEALPix’s unique latitude-dependent pixel alignment, an adjoint method for gradient computation in backpropagation, and out-of-core memory-efficient optimization for Legendre transforms. This approach makes \texttt{cuHPX} a powerful tool for high-resolution, large-scale spherical data processing across applications in climate science, astrophysics, and scientific AI.

\texttt{cuHPX} achieves high accuracy and efficiency in spherical harmonic transforms on HEALPix grids. By employing ring quadrature weights, it attains rapid spectral convergence across resolutions. Additionally, the differentiability of \texttt{cuHPX} supports integration into deep learning frameworks, with gradients analytically computed via adjoint methods for backpropagation. It also enables flexible regridding between HEALPix, equiangular, and other spherical grids, supporting both upscaling and downscaling. In performance comparisons, \texttt{cuHPX} delivers order-of-magnitude speedups over existing CPU-based libraries, underscoring its computational efficiency for large-scale applications.

\section{Acknowledgments}

This work was completed as part of Xiaopo Cheng's internship at NVIDIA. The authors gratefully acknowledge Dominique LaSalle, Murat Guney,  and Minseok Lee for helpful discussion and support.

\bibliography{references}
\bibliographystyle{unsrt}

\end{document}